\documentclass{article}
\usepackage{graphicx}
\usepackage{amsmath}
\usepackage{amsfonts}
\usepackage{amssymb}
\newtheorem{theorem}{Theorem}
\newtheorem{acknowledgement}[theorem]{Acknowledgement}

\newtheorem{corollary}[theorem]{Corollary}

\newtheorem{definition}[theorem]{Definition}

\newtheorem{lemma}[theorem]{Lemma}

\newtheorem{remark}[theorem]{Remark}

\newenvironment{proof}[1][Proof]{\textbf{#1.} }{\ \rule{0.5em}{0.5em}}

\begin{document}

\author{Jon E. Tyson\thanks{jonetyson@post.harvard.edu}
\and Jefferson Lab, Harvard University, Cambridge, MA 02138 USA}
\title{Operator-Schmidt decompositions\\and the Fourier transform,\\with applications to the operator-Schmidt numbers of unitaries}
\date{July 1, 2003}
\maketitle
\begin{abstract}
\noindent The operator-Schmidt decomposition is useful in quantum information
theory for quantifying the nonlocality of bipartite unitary operations. We
construct a family of unitary operators on $\mathbb{C}^{n}\otimes
\mathbb{C}^{n}$ whose operator-Schmidt decompositions are computed using the
discrete Fourier transform. As a corollary, we produce unitaries on
$\mathbb{C}^{3}\otimes\mathbb{C}^{3}$ with operator-Schmidt number $S$ for
every $S\in\left\{  1,...,9\right\}  $. This corollary was unexpected, since
it contradicted reasonable conjectures of Nielsen et al [Phys. Rev. A
\textbf{67} (2003) 052301] based on intuition from a striking result in the
two-qubit case. By the results of D\"{u}r, Vidal, and Cirac [Phys. Rev. Lett.
\textbf{89} (2002) 057901], who also considered the two-qubit case, our result
implies that there are nine equivalence classes of unitaries on $\mathbb{C}%
^{3}\otimes\mathbb{C}^{3}$ which are probabilistically interconvertible by
(stochastic) local operations and classical communication. As another
corollary, a prescription is produced for constructing maximally-entangled
unitaries from biunimodular functions. Reversing tact, we state a generalized
operator-Schmidt decomposition of the quantum Fourier transform considered as
an operator $\mathbb{C}^{M_{1}}\otimes\mathbb{C}^{M_{2}}\rightarrow
\mathbb{C}^{N_{1}}\otimes\mathbb{C}^{N_{2}}$, with $M_{1}M_{2}=N_{1}N_{2}$.
This decomposition shows (by Nielsen's bound) that the communication cost of
the QFT remains maximal when a net transfer of qudits is permitted. In an
appendix, a canonical procedure is given for removing basis-dependence for
results and proofs depending on the ``magic basis'' introduced in [S. Hill and
W. Wootters, ``Entanglement of a pair of quantum bits,'' Phys Rev. Lett
\textbf{78} (1997) 5022-5025].

\end{abstract}

\section{Introduction}

This paper addresses some open problems (questions \ref{firstproblem}%
-\ref{lastproblem} below) concerning the operator-Schmidt decomposition
\cite{NielsenThesis} (see definition \ref{genopschmdef} below), which is
useful in quantum information theory \cite{Preskill Notes} \cite{nielsenbook}
for quantifying nonlocality of bipartite unitary operations. Our main results
are obtained by constructing a family of unitaries on $\mathbb{C}^{N}%
\otimes\mathbb{C}^{N}$ with computable operator-Schmidt decompositions, a
result which should facilitate further study of this decomposition.

D\"{u}r, Vidal, and Cirac \cite{Dur Vidal Cirac} used operator-Schmidt numbers
to determine when there exists a probabilistic\footnote{i.e. succeeding with a
nonzero probability} simulation of a unitary $\tilde{U}$ on $\mathbb{C}%
^{d}\otimes\mathbb{C}^{d}$ with a \textit{single} application of a given
unitary $U$ on $\mathbb{C}^{d}\otimes\mathbb{C}^{d}$ aided by
(stochastic)\ local operations, classical communication, and ancilla. In
particular, they show that this simulation can occur iff $\operatorname{Sch}%
\left(  U\right)  \geq\operatorname{Sch}\left(  \tilde{U}\right)  $, where
$\operatorname{Sch}\left(  U\right)  $ is the number of nonzero
operator-Schmidt coefficients of $U$ (see definition \ref{genopschmdef}
below).\footnote{D\"{u}r et al. note that, for example, entanglement
purification is a probabilistic process, so it is natural to consider
probabilistic simulation of its component gates.}

Intriguingly, D\"{u}r, Vidal, and Cirac observed that a unitary acting on two
qubits may have operator-Schmidt number $1$, $2$, or $4$, but not
$3$.\footnote{This fact was rediscovered by Nielsen et al.
\cite{DynamicalStrength}.} Thus there exist three equivalence classes of two
qubit unitary operations under probabilistic local interconversion [using
(S)LOCC], with the successive classes represented by the identity,
$\operatorname{CNOT}$, and $\operatorname{SWAP}$ operation, respectively.
Their observation followed immediately from the canonical decomposition of
two-qubit unitaries \cite{Khaneja Brockett Glaser}:\footnote{Kraus and Cirac
have a constructive ``magic basis'' proof \cite{Kraus and Cirac}. The
invariants of this decomposition were first discovered by Makhlin
\cite{makhlin}.} any two-qubit unitary operation $U_{AB}\in\operatorname{SU}%
\left(  4\right)  $ can be written in the following standard form
\begin{equation}
U_{AB}=\left(  V_{A}\otimes W_{B}\right)  \exp\left(  i\sum_{k=1}^{3}\mu
_{k}\sigma_{k}^{A}\otimes\sigma_{k}^{B}\right)  \left(  \tilde{V}_{A}%
\otimes\tilde{W}_{A}\right)  ,\label{canonical decomposition}%
\end{equation}
where $V_{A}$, $W_{B}$, $\tilde{V}_{A}$, and $\tilde{W}_{B}$ are local
unitaries and where the $\sigma_{k}$ are the Pauli operators with $\sigma
_{0}\equiv1$, and
\[
\pi/4\geq\mu_{1}\geq\mu_{2}\geq\left|  \mu_{3}\right|  \geq0\text{.}%
\]
(Since the Schmidt coefficients of $U_{AB}$ are unaffected by the local
$V_{A}$, ..., $\tilde{W}_{B}$, their claim reduced to a simple calculation of
the operator-Schmidt coefficients of the exponential.)

An interesting problem posed by Nielsen et al \cite{DynamicalStrength} is to
find the allowed operator-Schmidt numbers of unitaries on $\mathbb{C}%
^{n}\otimes\mathbb{C}^{m}$. Since there is no known generalization of the
canonical decomposition $\left(  \ref{canonical decomposition}\right)  $ to
unitaries on $\mathbb{C}^{n}\otimes\mathbb{C}^{m}$ for $\max\left(
n,m\right)  >2$,\footnote{An interesting restriction of this open problem is
to illuminate the nonlocal structure of maximally-entangled bipartite
unitaries. (See definition \ref{def of maximally entangled operators}.)} at
present a different method is required to solve this problem. (The $n=m=3$
case is solved below.)

The operator-Schmidt decomposition was introduced by Nielsen
\cite{NielsenThesis} in consideration of the following problem of coherent
communication complexity:

\begin{quotation}
\noindent Suppose Alice has $n_{a}$ qubits and Bob has $n_{b}$ qubits, and
they wish to perform some general unitary operation $U$ on their $n_{a}+n_{b}
$ qubits. How many qubits of quantum communication are required to achieve
this goal?
\end{quotation}

\noindent Nielsen proved that the minimum number $Q_{0}\left(  U\right)  $ of
such qubits satisfies the following bound \cite{nielsentalk}:
\begin{equation}
K_{\text{har}}\left(  U\right)  \leq Q_{0}\left(  U\right)  \leq2\min\left(
n_{a},n_{b}\right)  \text{,\label{Nielsens Qubit bound}}%
\end{equation}
where the \textit{Hartley strength }$K_{\text{har}}$ satisfies
\[
K_{\text{har}}\left(  U\right)  =\log_{2}\left(  \operatorname{Sch}\left(
U\right)  \right)  ,
\]
where $\operatorname{Sch}\left(  U\right)  $, defined in definition
\ref{genopschmdef} below, is the number of nonzero operator-Schmidt
coefficients of $U$. It was assumed that Alice and Bob have the use of
ancilla, but they must separately retain their (modified) data qubits at the
end of the computation. The upper bound of $\left(  \text{\ref{Nielsens Qubit
bound}}\right)  $ is trivial, for Alice could simply send all her bits to Bob
and let him send them back, or vice-versa. We emphasize that the communication
complexity $Q_{0}\left(  U\right)  $ is the communication cost of exact
computation of one application of $U$. An interesting open problem is to
consider the communication cost of approximate computation of $U^{\otimes M}$,
where the error goes to zero in some appropriate sense for large
$M$.\footnote{It would be very interesting to know if this assymptotic cost
for approximate computation depends only on the operator-Schmidt coefficients
of $U$. The reader is warned that the entanglement $K_{\operatorname{Sch}%
}\left(  U\right)  $ of $U:\mathcal{A}\otimes\mathcal{B}\rightarrow
\mathcal{A}\otimes\mathcal{B}$ considered as an element of the vector-space
$B\left(  \mathcal{A}\right)  \otimes B\left(  \mathcal{B}\right)  $ (see
definition \ref{def of maximally entangled operators}) was shown by Nielsen et
al \cite{DynamicalStrength} not to satisfy the chaining property. In
particular, there exist $U,V$ such that $K_{\operatorname{Sch}}\left(
UV\right)  >K_{\operatorname{Sch}}\left(  U\right)  +K_{\operatorname{Sch}%
}\left(  V\right)  $.}

Nielsen applied his abstract bound to show that the communication complexity
of the quantum Fourier transform is maximal, first in the case of $n_{a}%
=n_{b}$ \cite{nielsentalk}\cite{NielsenThesis}, and then (with collaborators)
in the case $n_{a}\leq n_{b}$ \cite{DynamicalStrength}, where Alice holds the
most-significant qubits.\footnote{See section \ref{genFourSchmDecomp section}
for a precise statement of this problem.} This was extended to $n_{a}>n_{b}$
and to arbitrary qudits in \cite{Tyson}. In section \ref{genFourSchmDecomp
section} we extend this result to the case that a net transfer of data qudits
is permitted.

\subsection{Results}

The main result of this paper is the construction of a family of unitaries on
$\mathbb{C}^{N}\otimes\mathbb{C}^{N}$ whose Schmidt decompositions are
computable using Fourier analysis. Specifically, Theorem \ref{smfcl} gives a
set $\left\{  \Phi_{\alpha\beta}\right\}  $ of vectors in a tensor product of
two Hilbert spaces of dimension $N$ such that the Schmidt-coefficients of the
diagonal operator
\begin{equation}
D=\sum_{\alpha,\beta=0}^{N-1}\lambda\left(  \alpha,\beta\right)  \,\left|
\Phi_{\alpha\beta}\right\rangle \left\langle \Phi_{\alpha\beta}\right|
\label{illdecomposethis}%
\end{equation}
are the nonzero values of $\left|  \hat{\lambda}\left(  \alpha,\beta\right)
\right|  $, where $\hat{\lambda}$ is the discrete Fourier transform.

Furthermore, this paper addresses the following questions concerning the
operator-Schmidt decomposition:

\begin{enumerate}
\item \label{firstproblem}What operator-Schmidt numbers $S$ occur in unitary
operators on $\mathbb{C}^{3}\otimes\mathbb{C}^{3}$?

\item  How can one construct maximally-entangled unitaries on $\mathbb{C}%
^{N}\otimes\mathbb{C}^{N}$?

\item \label{lastproblem}Can one generalize the results of \cite{nielsentalk},
\cite{NielsenThesis}, \cite{DynamicalStrength}, and \cite{Tyson} to show that
the communication cost of quantum Fourier transform for data shared between
two parties remains maximal if a net transfer of data qudits is allowed to occur?
\end{enumerate}

The cases $S\notin\left\{  2,4\right\}  $ of question $1$ are resolved using
$\left(  \ref{illdecomposethis}\right)  $ by considering the cardinalities of
the support of Fourier transforms of phase-valued functions $\lambda\left(
\alpha,\beta\right)  =\exp\left(  i\theta_{\alpha\beta}\right)  $. Resolving
the remaining cases $S\in\left\{  2,4\right\}  $ by inspection, Theorem
\ref{Theorem Schmidt c3 ten c3} shows that there exists unitaries on
$\mathbb{C}^{3}\otimes\mathbb{C}^{3}$ with arbitrary Schmidt number
$S\in\left\{  1,...,9\right\}  $. By the work of D\"{u}r, Vidal, and Cirac
\cite{Dur Vidal Cirac}, this result implies that there are nine equivalence
classes of unitaries on $\mathbb{C}^{3}\otimes\mathbb{C}^{3}$ which are
probabilistically interconvertible by (stochastic) local operations and
classical communication.

Using the diagonal operator $\left(  \ref{illdecomposethis}\right)  $,
question 2 is partially answered by existing mathematical studies of
\textit{biunimodular functions}, that is phase-valued functions whose discrete
Fourier transforms are also phase-valued. Using a construction of Bj\"{o}rck
and Saffari \cite{Bjorck and Saffari}, uncountably-many maximally entangled
unitaries on $\mathbb{C}^{N}\otimes\mathbb{C}^{N}$ may be constructed for $N $
divisible by a square. However, it remains to check whether any two of the
constructed $A,$ $B$ are inequivalent in the sense that
\begin{equation}
A=\left(  U\otimes W\right)  B\left(  X\otimes Y\right)  \text{,\label{before
and after local unitaries}}%
\end{equation}
for local unitaries $U,$ $W,$ $X,$ and $Y$. The problem of how to verify such
equivalences completely has not (to our knowledge) been worked out for general
$A$ and $B$, and is left as an open problem. (This is related to \cite{Grassl
Rotteler Beth}, however.)

Question 3 is answered by computing the generalized operator-Schmidt
decomposition of the quantum Fourier transform as a map from $\mathbb{C}%
^{M_{1}}\otimes\mathbb{C}^{M_{2}}$ to $\mathbb{C}^{N_{1}}\otimes
\mathbb{C}^{N_{2}}$ and applying a slight modification of Nielsen's bound.

In the appendix we remark on the ``magic basis'' of Hill and Wootters
\cite{Hill and Wootters}.

\subsection{Definitions and Notation}

\begin{definition}
\label{genopschmdef}Let $\mathcal{A}$, $\mathcal{A}^{\prime}$, $\mathcal{B}$,
and $\mathcal{B}^{\prime}$ be finite-dimensional Hilbert spaces, and let
$F:\mathcal{A}\otimes\mathcal{B}\rightarrow\mathcal{A}^{\prime}\otimes
\mathcal{B}^{\prime}$ be a nonzero linear transformation. The
\textbf{Hilbert-Schmidt space} $B\left(  \mathcal{A}\rightarrow\mathcal{A}%
^{\prime}\right)  $ is the Hilbert space of linear transformations from
$\mathcal{A}$ to $\mathcal{A}^{\prime}$ under the Hilbert-Schmidt inner
product
\[
\left\langle C,D\right\rangle _{B\left(  \mathcal{A}\rightarrow\mathcal{A}%
^{\prime}\right)  }=\operatorname*{Tr}_{\mathcal{A}}C^{\dag}D,
\]
where
\[
\left\langle C^{\dag}\psi,\phi\right\rangle _{\mathcal{A}}=\left\langle
\psi,C\phi\right\rangle _{\mathcal{A}^{\prime}}%
\]
for all $\phi\in\mathcal{A}$ and $\psi\in\mathcal{A}^{\prime}$. For
simplicity, we define $B\left(  \mathcal{H}\right)  =B\left(  \mathcal{H}%
\rightarrow\mathcal{H}\right)  $. A \textbf{generalized operator-Schmidt
decomposition} of $F$ is a decomposition of the form
\begin{equation}
F=\sum_{k=1}^{\operatorname{Sch}\left(  F\right)  }\lambda_{k}\;A_{k}\otimes
B_{k}\text{, \ }\lambda_{k}>0,\label{genschmidtdecomp}%
\end{equation}
where the $\left\{  A_{k}\right\}  _{k=1...\operatorname{Sch}\left(  F\right)
}$ and $\left\{  B_{k}\right\}  _{k=1...\operatorname{Sch}\left(  F\right)  }$
are orthonormal subsets of $B\left(  \mathcal{A}\rightarrow\mathcal{A}%
^{\prime}\right)  $ and $B\left(  \mathcal{B}\rightarrow\mathcal{B}^{\prime
}\right)  $, respectively.\footnote{But not necessarily bases.} The quantity
$\operatorname{Sch}\left(  F\right)  $ is the \textbf{Schmidt number}, and the
$\lambda_{k}$ are the Schmidt coefficients. Equation $\left(
\ref{genschmidtdecomp}\right)  $ is an \textbf{operator-Schmidt decomposition}
when restricted to the special case $\mathcal{A}=\mathcal{A}^{\prime}$ and
$\mathcal{B}=\mathcal{B}^{\prime}$.
\end{definition}

We remark that the generalized operator-Schmidt decomposition is just a
special case of the well-known Schmidt decomposition \cite{nielsenbook}
\begin{equation}
\psi=\sum_{k=1}^{\operatorname{Sch}\left(  \psi\right)  }\lambda_{k}%
\;e_{k}\otimes f_{k}\text{, \ }\lambda_{k}>0\label{vectorschmidt}%
\end{equation}
of a vector $\psi\in\mathcal{H}\otimes\mathcal{K}$ $,$ where the $\left\{
e_{k}\right\}  $ and $\left\{  f_{k}\right\}  $ are orthonormal. In
particular, one sets $\mathcal{H}=B\left(  \mathcal{A}\rightarrow
\mathcal{A}^{\prime}\right)  ,$ $\mathcal{K}=B\left(  \mathcal{B}%
\rightarrow\mathcal{B}^{\prime}\right)  ,$ and $\psi=F\in B\left(
\mathcal{A}\otimes\mathcal{B}\rightarrow\mathcal{A}^{\prime}\otimes
\mathcal{B}^{\prime}\right)  $. The decomposition $\left(
\ref{genschmidtdecomp}\right)  $ is then obtained by identifying $B\left(
\mathcal{A}\rightarrow\mathcal{A}^{\prime}\right)  \otimes B\left(
\mathcal{B}\rightarrow\mathcal{B}^{\prime}\right)  $ with $B\left(
\mathcal{A}\otimes\mathcal{B}\rightarrow\mathcal{A}^{\prime}\otimes
\mathcal{B}^{\prime}\right)  $ under the natural isomorphism.\footnote{In
particular, there exists a unique unitary $\Xi:B\left(  \mathcal{A}%
\rightarrow\mathcal{A}^{\prime}\right)  \otimes B\left(  \mathcal{B}%
\rightarrow\mathcal{B}^{\prime}\right)  \rightarrow B\left(  \mathcal{A}%
\otimes\mathcal{B}\rightarrow\mathcal{A}^{\prime}\otimes\mathcal{B}^{\prime
}\right)  $ usch that $\left(  \Xi\left(  A\tilde{\otimes}B\right)  \right)
\left(  f\otimes g\right)  =\left(  Af\right)  \otimes\left(  Bg\right)  $ for
all $f\in\mathcal{A}$ and $g\in\mathcal{B}$. Here $\tilde{\otimes}$ denotes
the defining formal tensor product of $B\left(  \mathcal{A}\rightarrow
\mathcal{A}^{\prime}\right)  \otimes B\left(  \mathcal{B}\rightarrow
\mathcal{B}^{\prime}\right)  $, considering the factors as abstract Hilbert
spaces.} Note that $\operatorname{Sch}\left(  \psi\right)  $ and the set
$\left\{  \lambda_{k}\right\}  _{k=1,...,\operatorname{Sch}\left(
\psi\right)  }$ are independent of the choice of decomposition, since they are
just the rank and set of singular values of the map $\left|  f\right\rangle
_{\mathcal{K}}\mapsto\left\langle \psi\right|  \;\left|  f\right\rangle
_{\mathcal{K}}:\mathcal{K}\rightarrow\mathcal{H}^{\ast}$,
respectively.\footnote{See definition $\ref{absnonhhstar}$ for the
Hilbert-space structure of the dual space $\mathcal{H}^{\ast}$.} Furthermore,

\begin{definition}
\label{def of maximally entangled operators}The \textbf{Schmidt strength}
$K_{\operatorname{Sch}}\left(  F\right)  \;$ of $F:\mathcal{A}\otimes
\mathcal{B}\rightarrow\mathcal{A}^{\prime}\otimes\mathcal{B}^{\prime}$
\cite{DynamicalStrength} is the entanglement of $F$ considered as an element
of $B\left(  \mathcal{A}\rightarrow\mathcal{A}^{\prime}\right)  \otimes
B\left(  \mathcal{B}\rightarrow\mathcal{B}^{\prime}\right)  $.\footnote{Hence
$K_{\operatorname{Sch}}\left(  F\right)  =S\left(  \operatorname*{Tr}%
_{B\left(  \mathcal{A}\rightarrow\mathcal{A}^{\prime}\right)  }\left|
F\right\rangle \left\langle F\right|  \right)  $, where $S$ is the von-Neuman
entropy $S\left(  \rho\right)  =-\operatorname*{Tr}\rho\log\rho$.} $F$ is said
to be \textbf{maximally-entangled} if $K_{\operatorname{Sch}}\left(  F\right)
$ is maximized or, equivalently, if $\operatorname{Sch}\left(  F\right)
=\min\left(  \dim\left(  \mathcal{A}\right)  \dim\left(  \mathcal{A}^{\prime
}\right)  ,\dim\left(  \mathcal{B}\right)  \dim\left(  \mathcal{B}^{\prime
}\right)  \right)  $ and all the operator-Schmidt coefficients are equal.
\end{definition}

We note that the Schmidt-number condition on maximally-entangled operators
implies that they have maximal communication cost by Nielsen's bound $\left(
\ref{Nielsens Qubit bound}\right)  $ (see also the slight modification,
(\ref{sdfasdfs}) below).

\section{\label{foursec}Schmidt decompositions given by the Fourier transform}

The goal of this section is to construct the family of diagonal operators
$\left(  \ref{illdecomposethis}\right)  $, whose operator-Schmidt coefficients
are computed using the discrete Fourier transform. There are two ingredients
in this construction:

\begin{enumerate}
\item  The well-known isomorphism between $\mathcal{H}\otimes\mathcal{H}%
^{\ast}$ (defined below) and $B\left(  \mathcal{H}\right)  $, which allows
application of the tools of operator theory to the study of bipartite Hilbert spaces.

\item  The characterization (up to a phase) of the discrete Fourier transform
by its action by conjugation on the Heisenberg-Weyl algebra.
\end{enumerate}

\begin{definition}
\label{absnonhhstar}Let $\mathcal{H}$ be a Hilbert space of dimension $N$ with
inner product\footnote{We take linear products to be linear in the second
argument.} $\left\langle \bullet,\bullet\right\rangle _{\mathcal{H}}$, and let
$\mathcal{H}^{\ast}$ be its dual.\footnote{The dual space $\mathcal{H}^{\ast}$
is the set of linear functionals $\ell:\mathcal{H}\rightarrow\mathbb{C}$. In
Dirac notation, $\mathcal{H}^{\ast}$ is the space of bras.} Define the natural
antilinear map $f\mapsto\bar{f}:\mathcal{H}\rightarrow\mathcal{H}^{\ast}$ by
\begin{equation}
\overline{\left|  \psi\right\rangle }=\left\langle \psi\right|
\label{firstbar}%
\end{equation}
and endow $\mathcal{H}^{\ast}$ with the inner product $\left\langle \bar
{f},\bar{g}\right\rangle _{\mathcal{H}^{\ast}}=\left\langle g,f\right\rangle
_{\mathcal{H}}$. For a linear operator $A:\mathcal{H}\rightarrow\mathcal{H}$,
define the \textbf{adjoint }$A^{\dag}:\mathcal{H}\rightarrow\mathcal{H}$ by
\[
\left\langle f,Ag\right\rangle _{\mathcal{H}}=\left\langle A^{\dag
}f,g\right\rangle _{\mathcal{H}}%
\]
and the \textbf{conjugate} $\bar{A}:\mathcal{H}^{\ast}\rightarrow
\mathcal{H}^{\ast}$ by\footnote{The suggestive use of bar-notation in $\left(
\ref{firstbar}\right)  -\left(  \ref{secondbar}\right)  $ is motivated by the
following formulas: $\psi=\sum_{k}a_{k}\left|  k\right\rangle \Rightarrow
\bar{\psi}=\sum_{k}\bar{a}_{k}\overline{\left|  k\right\rangle }$ and
$A\left|  j\right\rangle =\sum_{k}a_{jk}\left|  k\right\rangle \Rightarrow
\bar{A}\,\overline{\left|  j\right\rangle }=\sum_{k}\bar{a}_{jk}%
\overline{\left|  k\right\rangle }$.\vspace{0.04in}}
\begin{equation}
\bar{A}\bar{f}=\overline{\,Af\,\text{.}}\label{secondbar}%
\end{equation}
The \textbf{natural isomorphism} $A\mapsto\left.  \left|  A\right\rangle
\right\rangle _{\mathcal{H}\otimes\mathcal{H}^{\ast}}:B\left(  \mathcal{H}%
\right)  \rightarrow\mathcal{H}\otimes\mathcal{H}^{\ast}$ is the unitary map
satisfying
\[
A=\left|  f\right\rangle \left\langle g\right|  \Longrightarrow\left.  \left|
A\right\rangle \right\rangle _{\mathcal{H}\otimes\mathcal{H}^{\ast}}%
=f\otimes\bar{g}\text{,}%
\]
for all $f,g\in\mathcal{H}$, where $\otimes$ on the right-hand-side is the
defining formal Hilbert-space tensor product of $\mathcal{H}\otimes
\mathcal{H}^{\ast}$.\footnote{The double-ket notation goes back to
\cite{Baranger}. Equivalently, $\left\langle f\otimes\bar{g}\right|
\cdot\left.  \left|  A\right\rangle \right\rangle _{\mathcal{H}\otimes
\mathcal{H}^{\ast}}=\left\langle f,Ag\right\rangle _{\mathcal{H}}\;$for all
$f,g\in\mathcal{H}$, where $\cdot$ is the inner product on $\mathcal{H}%
\otimes\mathcal{H}^{\ast}$.} Let $\mathbb{Z}_{N}=\left\{  0,...,N-1\right\}  $
and $\mathbb{Z}_{N}^{2}=\mathbb{Z}_{N}\times\mathbb{Z}_{N}$. The computational
basis is denoted by $\left\{  \left|  j\right\rangle \right\}  _{j\in
\mathbb{Z}_{N}}\subseteq\mathcal{H}$.
\end{definition}

The following lemma is a basis-free version of equations 6 and 10 of
\cite{DAriano Presti Sacchi}, with a similar proof:

\begin{lemma}
\label{simplesqueeze}Let $A,B,C\in B\left(  \mathcal{H}\right)  $. Then
$\left(  A\otimes\bar{B}\right)  \,\left.  \left|  C\right\rangle
\right\rangle _{\mathcal{H}\otimes\mathcal{H}^{\ast}}=\left.  \left|
ACB^{\dag}\right\rangle \right\rangle _{\mathcal{H}\otimes\mathcal{H}^{\ast}}%
$. Furthermore, $\left.  \left|  C\right\rangle \right\rangle _{\mathcal{H}%
\otimes\mathcal{H}^{\ast}}$ is maximally entangled iff $C$ is a nonzero scalar
multiple of a unitary.
\end{lemma}

The second ingredient in our construction is the following

\begin{theorem}
[H. Weyl]Let $\mathcal{H}$ and $N$ be as in definition \ref{absnonhhstar}.
Then any irreducible unitary representation of the group generated by the
discrete Weyl relations
\begin{align}
R^{n}  & =I\text{ iff }n\in N\mathbb{Z}\\
T^{n}  & =I\text{ iff }n\in N\mathbb{Z}\\
RT  & =\exp\left(  -\frac{2\pi i}{N}\right)  TR\text{.\label{weylrelation}}%
\end{align}
is unitarily equivalent to one in which $R$ and $T$ are represented on
$\mathcal{H}$ as the right-shift operator and twist operator, respectively:
\begin{align}
R\left|  j\right\rangle  & =\left|  j+1\,\operatorname{mod}\,N\right\rangle
\text{, }j\in\mathbb{Z}_{N}\label{rsdef}\\
T\left|  j\right\rangle  & =\exp\left(  \frac{2\pi ij}{N}\right)  \left|
j\right\rangle \text{.\label{twistdef}}%
\end{align}
Furthermore, if $F$ satisfies the associated Fourier relations
\begin{align*}
FRF^{-1}  & =T\\
FTF^{-1}  & =R^{-1}%
\end{align*}
then $F$ will be simultaneously represented (up to a scalar factor $\lambda$)
as the discrete Fourier transform:
\[
\left\langle j\right|  F\left|  k\right\rangle =\frac{\lambda}{\sqrt{N}}%
\exp\left(  \frac{2\pi i}{N}jk\right)  \text{.}%
\]
\end{theorem}

The first part of the theorem is given in \cite{weylqgroup}. The second part
follows trivially from Schur's lemma. Weyl considered the representations of
the discrete Weyl relations because they are a finite-dimensional analogue of
the canonical commutation relation $\left[  P,Q\right]  =-i$ for self-adjoint
$P$ and $Q$.\footnote{See \cite{reedsimon3} for the representations of the
infinite-dimensional Weyl relations (due to von Neumann). See
\cite{reedsimon1} for their relationship to the CCR, and for an example
(essentally due to Ed Nelson) of an irreducible representation of the CCR on
$L^{2}\left(  \mathbb{R}\right)  $ that is \textit{not} unitarily equivalent
to $Q=x$, $P=-i$\thinspace$d/dx$.}

\begin{definition}
The discrete Fourier transform of functions on $\mathbb{Z}_{N}^{2}$ is given
by
\[
\hat{\lambda}\left(  a,b\right)  =\frac{1}{N}\sum_{\alpha,\beta=0}^{N-1}%
\exp\left(  \frac{2\pi i}{N}\left(  \alpha a+\beta b\right)  \right)
\lambda\left(  \alpha,\beta\right)  \text{.}%
\]
\newpage
\end{definition}

\begin{theorem}
\label{smfcl}Take $R,T\in B\left(  \mathcal{H}\right)  $ to be given by
$\left(  \ref{rsdef}\right)  -\left(  \ref{twistdef}\right)  $. Let
\[
\Phi_{\alpha\beta}=N^{-1/2}\left.  \left|  T^{\alpha}R^{-\beta}\right\rangle
\right\rangle _{\mathcal{H}\otimes\mathcal{H}^{\ast}}%
\]
for $\alpha,\beta\in\mathbb{Z}_{N}$. Then the $\Phi_{\alpha\beta}$ form a
maximally entangled orthonormal basis of $\mathcal{H}\otimes\mathcal{H}^{\ast
}$. Furthermore, for an arbitrary function $\lambda:\mathbb{Z}_{N}%
^{2}\rightarrow\mathbb{C}$, the diagonal operator $D$
\begin{equation}
D=\sum_{\alpha,\beta=0}^{N-1}\lambda\left(  \alpha,\beta\right)  \,\left|
\Phi_{\alpha\beta}\right\rangle \left\langle \Phi_{\alpha\beta}\right|
:\mathcal{H}\otimes\mathcal{H}^{\ast}\rightarrow\mathcal{H}\otimes
\mathcal{H}^{\ast}\text{,\label{nicediag}}%
\end{equation}
satisfies the relation
\begin{equation}
D=\frac{1}{N}\sum_{a,b=0}^{N-1}\hat{\lambda}\left(  a,b\right)  \times\left(
R^{a}T^{b}\right)  \otimes\overline{\left(  R^{a}T^{b}\right)  }\text{.}%
\end{equation}
In particular, a Schmidt decomposition of $D$ is given by
\begin{equation}
D=\sum_{a,b}\left|  \hat{\lambda}\left(  a,b\right)  \right|  \times\left(
\frac{\hat{\lambda}\left(  a,b\right)  }{\left|  \hat{\lambda}\left(
a,b\right)  \right|  }\frac{1}{\sqrt{N}}R^{a}T^{b}\right)  \otimes
\overline{\left(  \frac{1}{\sqrt{N}}R^{a}T^{b}\right)  }\text{,}%
\label{yuckyabsolutevalue}%
\end{equation}
where the summation is over the $a,b\in\mathbb{Z}_{N}$ such that $\hat
{\lambda}\left(  a,b\right)  \neq0$.
\end{theorem}

\begin{proof}
It was observed by Schwinger \cite{schwingerunitarybases} that the set
$\left\{  N^{-1/2}T^{\alpha}R^{-\beta}\right\}  _{\alpha,\beta\in
\mathbb{Z}_{N}}$ is an orthonormal basis of $B\left(  \mathcal{H}\right)  $.
That the $\Phi_{\alpha\beta}$ form an orthonormal basis of $\mathcal{H}%
\otimes\mathcal{H}^{\ast}$ follows by the natural isomorphism. Maximal
entanglement follows from the second part of lemma \ref{simplesqueeze}.

By lemma \ref{simplesqueeze} and the Weyl relations $\left(
\ref{weylrelation}\right)  $, each $\Phi_{\alpha\beta}$ is an eigenvector of
each $\left(  R^{a}T^{b}\right)  \otimes\overline{\left(  R^{a}T^{b}\right)
}$:
\begin{align}
\left(  R^{a}T^{b}\right)  \otimes\overline{\left(  R^{a}T^{b}\right)  }%
\;\Phi_{\alpha\beta}  & =N^{-1/2}\left.  \left|  R^{a}T^{b}T^{\alpha}%
R^{-\beta}\left(  R^{a}T^{b}\right)  ^{\dag}\right\rangle \right\rangle
\nonumber\\
& =\exp\left(  -\frac{2\pi i}{N}\left(  a\alpha+b\beta\right)  \right)
\;\Phi_{\alpha\beta}\text{.}\label{niceevaleqn}%
\end{align}
Since the $\Phi_{\alpha\beta}$ form an orthonormal basis, $\left(
\ref{niceevaleqn}\right)  $ becomes
\[
\left(  R^{a}T^{b}\right)  \otimes\overline{\left(  R^{a}T^{b}\right)  }%
=\sum_{\alpha\beta=0}^{N-1}\exp\left(  -\frac{2\pi i}{N}\left(  a\alpha
+b\beta\right)  \right)  \;\left|  \Phi_{\alpha\beta}\right\rangle
\left\langle \Phi_{\alpha\beta}\right|  \text{.}%
\]
By the Fourier inversion theorem,
\[
\left|  \Phi_{\alpha\beta}\right\rangle \left\langle \Phi_{\alpha\beta
}\right|  =\frac{1}{N^{2}}\sum_{a,b=0}^{N-1}\exp\left(  \frac{2\pi i}%
{N}\left(  a\alpha+b\beta\right)  \right)  \,\left(  R^{a}T^{b}\right)
\otimes\overline{\left(  R^{a}T^{b}\right)  }\text{.}%
\]
Hence
\begin{align*}
D  & =\sum_{\alpha,\beta=0}^{N-1}\lambda\left(  \alpha,\beta\right)  \;\left|
\Phi_{\alpha\beta}\right\rangle \left\langle \Phi_{\alpha\beta}\right| \\
& =\sum_{\alpha,\beta=0}^{N-1}\lambda\left(  \alpha,\beta\right)  \frac
{1}{N^{2}}\sum_{a,b=0}^{N-1}\exp\left(  \frac{2\pi i}{N}\left(  a\alpha
+b\beta\right)  \right)  \,\left(  R^{a}T^{b}\right)  \otimes\overline{\left(
R^{a}T^{b}\right)  }\\
& =\frac{1}{N}\sum_{a,b=0}^{N-1}\hat{\lambda}\left(  a,b\right)  \,\left(
R^{a}T^{b}\right)  \otimes\overline{\left(  R^{a}T^{b}\right)  }.
\end{align*}
By the orthonormality of the $N^{-1/2}R^{a}T^{b}$, $\left(
\ref{yuckyabsolutevalue}\right)  $ is a Schmidt decomposition.
\end{proof}

\begin{remark}
By the lemmas used in \cite{Kraus and Cirac} to prove the canonical
decomposition $\left(  \ref{canonical decomposition}\right)  $ one has the
following fact: Up to local unitaries in the sense of $\left(  \ref{before and
after local unitaries}\right)  $, for $N=2$ every unitary on $\mathcal{H}%
\otimes\mathcal{H}^{\ast}$ is of the form $\left(  \ref{nicediag}\right)  ,$
even if the $\Phi_{\alpha\beta}$ are replaced by an arbitrary
maximally-entangled basis.
\end{remark}

\section{\label{c3c3sec}Application to Schmidt numbers of unitaries.}

It this section we produce the allowed Schmidt numbers of unitaries on
$\mathbb{C}^{3}\otimes\mathbb{C}^{3}$, solving a special case of the problem
of Nielsen et al \cite{DynamicalStrength} which prompted our investigations
here. By Theorem \ref{smfcl}, one may produce a unitary of Schmidt number $S$
from a ``unimodular'' function $\lambda:\mathbb{Z}_{N}^{2}\rightarrow\left\{
\left|  z\right|  =1\right\}  $ whose Fourier transform $\hat{\lambda}$ has
support of cardinality $S$.

\begin{lemma}
\label{hardfourrest3}There exists a function $\lambda:\mathbb{Z}_{3}%
^{2}\rightarrow\left\{  \left|  z\right|  =1\right\}  $ such that the support
of $\hat{\lambda}$ has cardinality $S$ iff $S\in\left\{
1,3,5,6,7,8,9\right\}  $.
\end{lemma}

\begin{proof}
Define $g_{1}$ and $g_{3}:\mathbb{Z}_{3}\rightarrow\left\{  \left|  z\right|
=1\right\}  $ by declaring $g_{1}=1$ identically and choosing and $g_{3}$ such
that $\operatorname*{supp}\hat{g}_{3}=\mathbb{Z}_{3}$. Then the support of the
Fourier transform of $\left(  g_{a}\otimes g_{b}\right)  \left(  j,k\right)
=g_{a}\left(  j\right)  g_{b}\left(  k\right)  $ has cardinality
$S=ab\in\left\{  1,3,9\right\}  $. For a function $\lambda:\mathbb{Z}_{3}%
^{2}\rightarrow\mathbb{C}$, let $\Gamma\lambda$ be the $3\times3$ matrix whose
$j,k$ entry is $\lambda\left(  j,k\right)  $, $j,k\in\mathbb{Z}_{3}$. Setting
\[
\omega=\exp\left(  \frac{2\pi i}{3}\right)  \text{,}%
\]
one has the following table of unimodular $\lambda_{S}$ such that the support
of $\hat{\lambda}_{S}$ has cardinality $S$:
\[%
\begin{tabular}
[c]{|c|c|c|}\hline
$S$ & $\Gamma\lambda_{S}$ & $\overset{\;}{\Gamma\hat{\lambda}_{S}}$\\\hline
$5$ & $%
\begin{array}
[c]{c}%
\\
\left[
\begin{array}
[c]{ccc}%
1 & -1 & 1\\
\omega & -\omega & \omega^{2}\\
\omega^{2} & -\omega^{2} & \omega
\end{array}
\right] \\
\;
\end{array}
$ & $%
\begin{array}
[c]{c}%
\\
\left[
\begin{array}
[c]{ccc}%
0 & 0 & 0\\
1 & \omega^{2} & \omega\\
0 & 1-\omega & 1-\omega^{2}%
\end{array}
\right] \\
\;
\end{array}
$\\\hline
$6$ & $%
\begin{array}
[c]{c}%
\\
\left[
\begin{array}
[c]{ccc}%
1 & 1 & 1\\
\omega & \omega & \omega^{2}\\
\omega^{2} & \omega^{2} & \omega
\end{array}
\right] \\
\;
\end{array}
$ & $%
\begin{array}
[c]{c}%
\\
\left[
\begin{array}
[c]{ccc}%
0 & 0 & 0\\
1 & \omega^{2} & \omega\\
2 & 1+\omega & 1+\omega^{2}%
\end{array}
\right] \\
\;
\end{array}
$\\\hline
$7$ & $%
\begin{array}
[c]{c}%
\\
\left[
\begin{array}
[c]{ccc}%
\quad1 & \quad\omega & \quad\omega^{2}\\
\quad1 & -\omega^{2} & \quad\omega\\
-1 & \quad\omega^{2} & -\omega
\end{array}
\right] \\
\;
\end{array}
$ & $%
\begin{array}
[c]{c}%
\\
\frac{1}{3}\left[
\begin{array}
[c]{ccc}%
0 & 0 & 3\\
-2+2\omega & \quad1+2\omega & \quad7+2\omega\\
\quad2-2\omega & -1-2\omega & -1-2\omega
\end{array}
\right] \\
\;
\end{array}
$\\\hline
$8$ & $%
\begin{array}
[c]{c}%
\\
\left[
\begin{array}
[c]{ccc}%
\quad\omega & \quad\omega & \quad\omega^{2}\\
\quad1 & -\omega^{2} & \quad\omega^{2}\\
-1 & \quad1 & -\omega
\end{array}
\right] \\
\;
\end{array}
$ & $%
\begin{array}
[c]{c}%
\\
\frac{1}{3}\left[
\begin{array}
[c]{ccc}%
0 & -3+3\omega & 3\\
-2+2\omega & \quad1+2\omega & \quad4+5\omega\\
-1+\omega & -1-2\omega & -1-2\omega
\end{array}
\right] \\
\;
\end{array}
$\\\hline
\end{tabular}
\]

Now let $P\subseteq\mathbb{Z}_{3}^{2}$ have cardinality $S=2$ or $4$. We claim
that there exists a nonzero $v\in\mathbb{Z}_{3}^{2}$ such that there exits a
unique $x\in P$ such that $x+v\operatorname{mod}3\mathbb{Z}^{2}\in P$. For
$S=2$ this fact is trivial. For $S=4$, by a modular translation and a
rotation, one can assume without loss of generality that the points $\left(
0,0\right)  $ and $\left(  0,1\right)  $ are in $P$. But then either $\left(
0,2\right)  \in P$ or there is another adjacent pair $\left(  s,t\right)  $
and $\left(  s,t+1\right)  \in P$. In either case a contradiction follows by inspection.

Now suppose $\lambda$ is unimodular and $\hat{\lambda}$ has cardinality $2$ or
$4$. Let $P$ be the support of $\hat{\lambda}$ and take $v$ and $x$ to be as
in the previous paragraph. Then
\begin{align*}
0  & =\delta_{v,0}=\widehat{\left(  \bar{\lambda}\lambda\right)  }\left(
-v\,\operatorname{mod}\,3\mathbb{Z}^{2}\right) \\
& =\frac{1}{N}\sum_{w\in\mathbb{Z}_{3}^{2}}\overline{\hat{\lambda}}\left(
v+w\,\operatorname{mod}\,3\mathbb{Z}^{2}\right)  \hat{\lambda}\left(  w\right)
\\
& =\frac{1}{N}\overline{\hat{\lambda}}\left(  v+x\,\operatorname{mod}%
\,N\mathbb{Z}^{2}\right)  \hat{\lambda}\left(  x\right)  \neq0\text{,}%
\end{align*}
yielding a contradiction.
\end{proof}

\newpage

\begin{theorem}
\label{Theorem Schmidt c3 ten c3}There exist unitary operators on
$\mathbb{C}^{3}\otimes\mathbb{C}^{3}$ with Schmidt number $S$, for every
$S\in\left\{  1,...,9\right\}  $.
\end{theorem}

\begin{proof}
By Theorem \ref{smfcl} and lemma \ref{hardfourrest3}, all that remains is to
check that there exist unitaries on $\mathbb{C}^{3}\otimes\mathbb{C}^{3}$ with
the Schmidt numbers $2$ and $4$. Setting
\[
P_{1}=\operatorname{diag}\left(  1,0,0\right)  \text{, \ \ \ }P_{2}%
=\operatorname{diag}\left(  0,1,1\right)  ,
\]
both of the following unitary operators have Schmidt number $2:$%
\begin{align*}
U  & =P_{1}\otimes R+P_{2}\otimes I\\
V  & =R\otimes P_{1}+I\otimes P_{2}\text{,}%
\end{align*}
where $R$ is given by $\left(  \ref{rsdef}\right)  $. Furthermore, their
product
\[
UV=P_{1}R\otimes RP_{1}+P_{1}\otimes RP_{2}+P_{2}R\otimes P_{1}+P_{2}\otimes
P_{2}%
\]
has Schmidt number $4$, since this is already a Schmidt decomposition, except
for normalizations.
\end{proof}

\section{\label{fourconsec}A connection between maximally entangled unitaries
and biunimodular functions}

Theorem \ref{smfcl} gives some insight into the problem of constructing
maximally-entangled unitaries on $\mathbb{C}^{N}\otimes\mathbb{C}^{N}$. The
best-known example of such a unitary is the $\operatorname{SWAP}$ operator
$f\otimes g\mapsto g\otimes f\;$on $\mathbb{C}^{N}\otimes\mathbb{C}^{N}$, with
Schmidt decomposition
\begin{equation}
\operatorname{SWAP}=\sum_{j=1}^{N^{2}}A_{j}\otimes A_{j}^{\dag}%
\text{,\label{swapschmidtdecomp}}%
\end{equation}
where $\left\{  A_{j}\right\}  _{j=1...N^{2}}$ is \textit{any} orthonormal
basis of $B\left(  \mathbb{C}^{N}\right)  $.\footnote{Since $\left\langle
A_{j}\otimes A_{k}^{\ast},\operatorname{SWAP}\right\rangle _{B\left(
\mathbb{C}^{n}\otimes\mathbb{C}^{n}\right)  }=\left\langle A_{j}%
,A_{k}\right\rangle _{B\left(  \mathbb{C}^{n}\otimes\mathbb{C}^{n}\right)
}=\delta_{jk}$, equation $\left(  \ref{swapschmidtdecomp}\right)  $ is just
the coordinate expansion of $\operatorname{SWAP}$ in the orthonormal basis
$\left\{  A_{j}\otimes A_{k}^{\ast}\right\}  $ of $B\left(  \mathbb{C}%
^{n}\otimes\mathbb{C}^{n}\right)  $.} Furthermore, corollary \ref{Gen Four
Schm Combinatorics Corollary}, below, shows that the quantum Fourier transform
$\mathcal{F}_{M_{1}M_{2}\rightarrow N_{1}N_{2}}:\mathbb{C}^{M_{1}}%
\otimes\mathbb{C}^{M_{2}}\rightarrow\mathbb{C}^{N_{1}}\otimes\mathbb{C}%
^{N_{2}}$ is maximally-entangled in many cases, including the case where only
one species of qudit is present.\footnote{Special cases of the general result
$\left(  \ref{whenmaximallyentangled}\right)  $ were given in
\cite{NielsenThesis}\cite{DynamicalStrength}\cite{Tyson}.}

Theorem \ref{smfcl} shows that the diagonal operator $D$ on $\mathbb{C}%
^{N}\otimes\mathbb{C}^{N}$ $\left(  \ref{nicediag}\right)  $ is maximally
entangled iff $\lambda:\mathbb{Z}_{N}^{2}\rightarrow\mathbb{C}$ is
\textit{biunimodular }\cite{Bjorck}, i.e. both $\lambda$ and $\hat{\lambda}$
have ranges lying in the circle $\left\{  \left|  z\right|  =1\right\}  $. To
characterize the biunimodular functions on $\mathbb{Z}_{N}^{2}$ is a
generalization of a studied problem of considerable difficulty: to
characterize the biunimodular functions on $\mathbb{Z}_{N}$.

Known examples of biunimodular functions on $\mathbb{Z}_{N}^{2}$ come as
tensor products $f\left(  x\right)  g\left(  y\right)  $ of biunimodular
functions $f$ and $g$ on $\mathbb{Z}_{N}$. The first examples of biunimodular
functions on $\mathbb{Z}_{N}$ were known to Gauss: for odd $N$ there are the
biunimodular Gaussians $g_{N,a,b}:\mathbb{Z}_{N}\rightarrow\mathbb{C}$, for
$a,b\in\mathbb{Z}_{N}$ with $a$ coprime to $N$, given by
\[
g_{N,a,b}\left(  k\right)  =\exp\left(  \frac{2\pi i}{N}\left(  ak^{2}%
+bk\right)  \right)  \text{,\ \ \ }%
\]
and for even $N$ one has
\[
g_{N}\left(  k\right)  =\exp\left(  \frac{2\pi i}{N}k^{2}\right)  \text{.}%
\]
For $N$ divisible by a square, these Gaussian examples are special cases of
the following theorem:

\begin{theorem}
[Bj\"{o}rck and Saffari \cite{Bjorck and Saffari}]Let $n^{2}$ be the largest
square dividing $N$. If $n>1$ then there exist infinitely many biunimodular
functions on $\mathbb{Z}_{N}$. In particular, setting $m=N/n$,

\begin{enumerate}
\item [Case 1:]Either $n$ is even or $m$ is odd. An infinite set of
biunimodular functions $f_{\tau,\mathcal{C},\rho}:\mathbb{Z}_{N}%
\rightarrow\mathbb{C}$ is given by
\begin{equation}
f_{\tau,\mathcal{C},\rho}\left(  k\right)  =c_{h}\rho^{r\tau\left(  h\right)
+nr\left(  r-1\right)  /2}\text{,\ \ \label{mostcasesbiunimodular}\ }%
\end{equation}
where $k$ has ``mixed-decimal'' expansion $k=nr+h$ (with $0\leq h<n$, $0\leq
r<m$), where $\mathcal{C}=\left(  c_{0},...,c_{n-1}\right)  $ is an arbitrary
unimodular sequence of length $n$, $\tau$ is any permutation of $\left\{
0,1,...,n-1\right\}  $, and $\rho$ is any primitive $m$th root of unity.

\item[Case 2:] $n$ is odd and $m$ is even. Each function $g_{\tau
,\mathcal{C},\rho}:\mathbb{Z}_{N}\rightarrow\mathbb{C}$ of the following form
is biunimodular:
\[
g_{\tau,\mathcal{C},\rho}\left(  k\right)  =z_{k\operatorname{mod}2}\times
f_{\tau,\mathcal{C},\rho}\left(  k\,\operatorname{mod}\left(  N/2\right)
\right)  ,
\]
where $z$ is the sequence $z=\left(  1,i\right)  $ and where $f_{\tau
,\mathcal{C},\rho}:\mathbb{Z}_{N/2}\rightarrow\mathbb{C}$ is a biunimodular
function generated using case $1$.
\end{enumerate}
\end{theorem}

\noindent For further results on biunimodular functions, see \cite{Bjorck},
\cite{Backelin and Froberg}, \cite{Bjorck and Froberg 1}, and \cite{Bjorck and
Froberg 2}.

\section{\label{genFourSchmDecomp section}Generalized Schmidt decomposition of
the quantum Fourier transform}

In this section, we consider the communication complexity of the bipartite
quantum Fourier transform when a net transfer of data is allowed to occur
between the two parties, generalizing the decompositions of
\cite{NielsenThesis}\cite{DynamicalStrength}\cite{Tyson}.

\begin{definition}
The quantum Fourier transform $\mathcal{F}_{M_{1}M_{2}\rightarrow N_{1}N_{2}%
}:\mathbb{C}^{M_{1}}\otimes\mathbb{C}^{M_{2}}\rightarrow\mathbb{C}^{N_{1}%
}\otimes\mathbb{C}^{N_{2}}$, with $N=N_{1}N_{2}=M_{1}M_{2}$, is the unitary
map satisfying
\[
_{N_{1}}\left\langle j\right|  \,_{N_{2}}\left\langle k\right|  \;\mathcal{F}%
_{M_{1}M_{2}\rightarrow N_{1}N_{2}}\;\left|  \ell\right\rangle _{M_{1}%
}\,\left|  m\right\rangle _{M_{2}}=\frac{1}{\sqrt{N}}\exp\left(  \frac{2\pi
i}{N}\left(  jN_{2}+k\right)  \left(  \ell M_{2}+m\right)  \right)  \text{.}%
\]
The \textbf{communication cost} of a unitary operation $U:\mathbb{C}^{M_{1}%
}\otimes\mathbb{C}^{M_{2}}\rightarrow\mathbb{C}^{N_{1}}\otimes\mathbb{C}%
^{N_{2}}$ is given by
\[
Q_{0}\left(  U\right)  =\min\sum_{d=2}^{\infty}N_{d}\log_{2}\left(  d\right)
\text{,}%
\]
where the minimum is over all protocols to compute $U$ using ancilla, local
operations, and the transmission of $N_{d}$ qudits of dimension $d,$ for
$d\geq2$. The communication cost of $U$ is said to be\textbf{\ maximal} if
$Q_{0}\left(  U\right)  =\log_{2}\min\left(  M_{1}N_{1},M_{2}N_{2}\right)  $.
\end{definition}

This is just the usual quantum Fourier transform, with the data shared by
Alice and Bob using mixed-decimals $\left|  \ell\right\rangle _{M_{1}}\left|
m\right\rangle _{M_{2}}\leftrightarrow\left|  \ell M_{2}+m\right\rangle _{N}$
before (and $\left|  j\right\rangle _{N_{1}}\left|  k\right\rangle _{N_{2}%
}\leftrightarrow\left|  jN_{2}+k\right\rangle _{N}$ after) the computation. We
note that the dimensions of Alice and Bob's local Hilbert spaces change upon
each communication of a qudit, although the product of the dimensions remains
constant. Furthermore, the trivial bound
\begin{equation}
Q_{0}\left(  U\right)  \leq\log_{2}\min\left(  M_{1}N_{1},M_{2}N_{2}\right)
,\label{trivial com bound allowing net data transfer}%
\end{equation}
for any $U:\mathbb{C}^{M_{1}}\otimes\mathbb{C}^{M_{2}}\rightarrow
\mathbb{C}^{N_{1}}\otimes\mathbb{C}^{N_{2}}$ follows from the fact that Alice
could send all her qudits to the Bob, who would perform the computation and
send back the required ones (or vice-versa).

We now state the generalized Schmidt decomposition of $\mathcal{F}_{M_{1}%
M_{2}\rightarrow N_{1}N_{2}}$. A proof and derivation are not included, as
they scarcely differ from those in $\cite{Tyson}$.

\begin{definition}
Let $\mathbb{Z}_{N}=\left\{  0,...,N-1\right\}  $. The equivalence classes of
$\mathbb{Z}_{N_{2}}\times\mathbb{Z}_{M_{2}}$ mod $\left(  M_{1},N_{1}\right)
$ consist of all sets of the form
\[
C=\left\{  \left.  \left(  a+M_{1}k_{1},b+M_{2}k_{2}\right)  \,\right|
\;a\in\mathbb{Z}_{N_{2}}\text{, }b\in\mathbb{Z}_{M_{2}}\text{, }k_{1},k_{2}%
\in\mathbb{Z}\right\}  \cap\left(  \mathbb{Z}_{N_{2}}\times\mathbb{Z}_{M_{2}%
}\right)
\]
where addition is NOT modular.
\end{definition}

\noindent Note that we do not consider equivalence classes of $\mathbb{Z}%
_{N_{2}}\times\mathbb{Z}_{M_{2}}$ mod $\left(  N_{1},M_{1}\right)  $: the
order of the $N$'s and $M$'s switches.

\begin{theorem}
Let $N=N_{1}N_{2}=M_{1}M_{2}.$ Then $\mathcal{F}_{M_{1}M_{2}\rightarrow
N_{1}N_{2}}$ has generalized Schmidt decomposition
\[
\mathcal{F}_{M_{1}M_{2}\rightarrow N_{1}N_{2}}=\sum_{C}\lambda_{c}%
\;A_{C}\otimes B_{C},
\]
where the summation is over equivalence classes $C$ of $\mathbb{Z}_{N_{2}%
}\times\mathbb{Z}_{M_{2}}$ mod $\left(  M_{1},N_{1}\right)  $, and where
\begin{align*}
\lambda_{C}  & =\sqrt{\frac{N_{1}M_{1}\operatorname{Card}\left(  C\right)
}{N}}\\
\left(  A_{C}\right)  _{jk}  & =\frac{1}{\sqrt{N_{1}M_{1}}}\exp\left[
\frac{2\pi i}{N}\left(  N_{2}M_{2}jk+M_{2}k\hat{s}+N_{2}j\hat{t}\right)
\right]  \text{, for }\left(  \hat{s},\hat{t}\right)  \in C\\
\left(  B_{C}\right)  _{jk}  & =\frac{1}{\sqrt{\operatorname{Card}\left(
C\right)  }}\times\left\{
\begin{array}
[c]{cc}%
\exp\left(  \frac{2\pi i}{N}jk\right)  & \text{if\ }\left(  j,k\right)  \in
C\\
0 & \text{otherwise}%
\end{array}
\right.  \text{,}%
\end{align*}
with $\operatorname{Card}\left(  C\right)  $ denoting the cardinality of $C$.
Note that the definition of $A_{C}$ is independent of the choice of $\left(
\hat{s},\hat{t}\right)  \in C$.
\end{theorem}

\begin{corollary}
\label{Gen Four Schm Combinatorics Corollary}In all cases
\begin{equation}
\operatorname{Sch}\left(  \mathcal{F}_{M_{1}M_{2}\rightarrow N_{1}N_{2}%
}\right)  =\min\left(  M_{1}N_{1},M_{2}N_{2}\right)
\text{.\label{schmnumgenQFT}}%
\end{equation}
In particular, the communication cost of $Q_{0}\left(  \mathcal{F}_{M_{1}%
M_{2}\rightarrow N_{1}N_{2}}\right)  $ is maximal in all cases. Furthermore,
$\mathcal{F}_{M_{1}M_{2}\rightarrow N_{1}N_{2}}$ is maximally-entangled iff
\begin{equation}%
\begin{tabular}
[c]{c}%
$\left(  M_{1}\text{ is a factor of }N_{2}\text{ or }M_{1}>N_{2}\right)  $\\
and\\
$\left(  N_{1}\text{ is a factor of }M_{2}\text{ or }N_{1}>M_{2}\right)  $.
\end{tabular}
\label{whenmaximallyentangled}%
\end{equation}
Otherwise $\mathcal{F}_{M_{1}M_{2}\rightarrow N_{1}N_{2}}$ has at most four
distinct Schmidt coefficients (of various multiplicities), taking values of
the form
\[
\sqrt{\frac{N_{1}M_{1}}{N}a_{\pm}b_{\pm}}\text{,}%
\]
where we ignore Schmidt coefficients stated as zero, and where
\[%
\begin{array}
[b]{ccc}%
a_{+}=\left\lceil N_{2}/M_{1}\right\rceil  & , & a_{-}=\left\lfloor
N_{2}/M_{1}\right\rfloor \\
b_{+}=\left\lceil M_{2}/N_{1}\right\rceil  & , & b_{-}=\left\lfloor
M_{2}/N_{1}\right\rfloor
\end{array}
.
\]
\end{corollary}

\begin{proof}
Equation $\left(  \ref{schmnumgenQFT}\right)  $ follows by a simple counting
argument. That the communication cost is maximal then follows from a slight
modification of the work of Nielsen et al in $\cite{DynamicalStrength},$ as
follows.\footnote{This idea was mentioned vaguely in footnote $10$ of
\cite{DynamicalStrength} and in footnotes $1$ and $6$ of \cite{Tyson}.}
Replacing the Schmidt decomposition by the generalized Schmidt decomposition
in the definition of the Hartley strength \cite{DynamicalStrength} and
replacing the $\operatorname{SWAP}$ operator in section $III.B.3$ of
\cite{DynamicalStrength} by communication operators\footnote{As stated in
\cite{Tyson}, the communication operator has generalized Schmidt-decomposition
$C=\sum_{k=1}^{d_{2}}\sqrt{d_{1}d_{3}}A_{k}\otimes B_{k}$, where $A_{k}%
=d_{1}^{-1/2}\sum_{i=1}^{d_{1}}\left|  i\right\rangle \left\langle ik\right|
:$ $\mathbb{C}^{d_{1}}\otimes\mathbb{C}^{d_{2}}\rightarrow\mathbb{C}^{d_{1}}$
and $B_{k}=d_{3}^{-1/2}\sum_{i=1}^{d_{3}}\left|  ki\right\rangle \left\langle
i\right|  :\mathbb{C}^{d_{3}}\rightarrow\mathbb{C}^{d_{2}}\otimes
\mathbb{C}^{d_{3}}$.}
\begin{align}
C:\left(  \mathbb{C}^{d_{1}}\otimes\mathbb{C}^{d_{2}}\right)  \otimes
\mathbb{C}^{d_{3}}  & \rightarrow\mathbb{C}^{d_{1}}\otimes\left(
\mathbb{C}^{d_{2}}\otimes\mathbb{C}^{d_{3}}\right) \nonumber\\
\left(  f\otimes g\right)  \otimes h  & \mapsto f\otimes\left(  g\otimes
h\right)  \text{,\label{communication operator def}}%
\end{align}
one immediately obtains the following version of Nielsen's bound $\left(
\ref{Nielsens Qubit bound}\right)  $:
\begin{equation}
K_{\text{har}}\left(  \mathcal{F}_{M_{1}M_{2}\rightarrow N_{1}N_{2}}\right)
\leq Q_{0}\left(  \mathcal{F}_{M_{1}M_{2}\rightarrow N_{1}N_{2}}\right)
\text{.\label{sdfasdfs}}%
\end{equation}
Hence the left-hand side of $\left(  \ref{sdfasdfs}\right)  $ equals the
right-hand-side of $\left(  \ref{trivial com bound allowing net data
transfer}\right)  $, proving the communication cost is maximal, as claimed.
The rest of this corollary is trivial.
\end{proof}

\newpage

\section{Appendix: The magic basis, without the basis.}

The natural isomorphism $A\mapsto\left.  \left|  A\right\rangle \right\rangle
:B\left(  \mathcal{H}\right)  \rightarrow\mathcal{H}\otimes\mathcal{H}^{\ast}$
has allowed us application of the tools of operator theory on $B\left(
\mathcal{H}\right)  $ to the study of bipartite tensor product spaces in a
natural manner. In this spirit we list below the properties of the gradient of
the determinant, which we will relate to ``conjugation in the magic basis'' of
Hill and Wootters $\cite{Hill and Wootters}$:

\begin{theorem}
The determinant is everywhere-differentiable on $B\left(  \mathcal{H}\right)
$. In particular, the determinant has a gradient $\mathcal{G}$ $:B\left(
\mathcal{H}\right)  \rightarrow B\left(  \mathcal{H}\right)  $ such that
\[
\frac{d}{dt}\det\left(  A\right)  =\left\langle \mathcal{G}\left(  A\right)
,\frac{dA}{dt}\right\rangle _{B\left(  \mathcal{H}\right)  }%
\]
for differentiable functions $A:\mathbb{R}\rightarrow B\left(  \mathcal{H}%
\right)  .$ Define the corresponding map $\frak{D}:\mathcal{H}\otimes
\mathcal{H}^{\ast}\rightarrow\mathcal{H}\otimes\mathcal{H}^{\ast}$ by
\[
\frak{D}\left.  \left|  A\right\rangle \right\rangle =\left.  \left|
\mathcal{G}\left(  A\right)  \right\rangle \right\rangle \text{.}%
\]
Taking $\lambda\in\mathbb{C}$, $A,B\in B\left(  \mathcal{H}\right)  $,
$\psi\in\mathcal{H}\otimes\mathcal{H}^{\ast},$ and $N=\dim\left(
\mathcal{H}\right)  $, the functions $\mathcal{G}$ and $\frak{D}$ have the
following properties:

\begin{enumerate}
\item $\mathcal{G}$ is the continuous extension of the map $A\mapsto\left(
\left(  \det A\right)  A^{-1}\right)  ^{\dag}$ from invertible $A$ to all $A$.

\item $\mathcal{G}\left(  AB\right)  =\mathcal{G}\left(  A\right)
\mathcal{G}\left(  B\right)  $ and $\mathcal{G}\left(  A^{\dag}\right)
=\left(  \mathcal{G}\left(  A\right)  \right)  ^{\dag}$. In particular,
$\mathcal{G}$ acts independently on the factors of the polar decomposition.

\item $\frak{D}\left(  \left(  A\otimes\bar{B}\right)  \psi\right)  =\left(
\mathcal{G}\left(  A\right)  \otimes\overline{\mathcal{G}\left(  B\right)
}\right)  \frak{D}\left(  \psi\right)  $. In particular, if $A$ and $B$ are
unitary then $\frak{D}\left(  \left(  A\otimes\bar{B}\right)  \psi\right)
=\left(  \det A^{\dag}B\right)  \left(  A\otimes\bar{B}\right)  \frak{D}%
\left(  \psi\right)  $.

\item \label{maxent prop}$\frak{D}\left(  \psi\right)  =\alpha\psi$ for some
$\alpha\in\mathbb{C}$ iff $\psi$ is maximally entangled or zero. Furthermore,
for $N\geq3$ the maximizers of $\left\|  \frak{D}\left(  \psi\right)
\right\|  _{\mathcal{H}\otimes\mathcal{H}^{\ast}}/\left\|  \psi\right\|
_{\mathcal{H}\otimes\mathcal{H}^{\ast}}$ are precisely the maximally entangled states.

\item \label{preconcur prop}Temporarily allowing Schmidt coefficients to
vanish, the product of the Schmidt coefficients of $\psi$ is given by
$N^{-1}\left|  \left\langle \psi,\frak{D}\psi\right\rangle _{\mathcal{H}%
\otimes\mathcal{H}^{\ast}}\right|  $.

\item  Furthermore, if $N=2$ then

\begin{enumerate}
\item \label{conj prop}$\mathcal{G}$ and $\frak{D}$ are conjugations, i.e.
antiunitary maps squaring to the identity.

\item \label{sep prop}$\psi$ is separable iff $\left\langle \psi,\frak{D}%
\psi\right\rangle =0$.

\item  Denote $\left|  i\bar{j}\right\rangle \equiv\left|  i\right\rangle
\otimes\overline{\left|  j\right\rangle }\in\mathcal{H}\otimes\mathcal{H}%
^{\ast}$. Then each of the following vectors are invariant under $\frak{D}$:
\begin{equation}
\left\{  \left|  0\bar{0}\right\rangle +\left|  1\bar{1}\right\rangle
,\;i\left|  0\bar{0}\right\rangle -i\left|  1\bar{1}\right\rangle ,\;i\left|
0\bar{1}\right\rangle +i\left|  1\bar{0}\right\rangle ,\;\left|  0\bar
{1}\right\rangle -\left|  1\bar{0}\right\rangle \right\}
\text{.\label{wootersbar}}%
\end{equation}
Furthermore, they form an orthonormal basis.

\item  If $A=e^{i\theta}UP$, where $U\in SU\left(  2\right)  $ and
$P=\operatorname{diag}\left(  \lambda_{1},\lambda_{2}\right)  $ is positive,
then $\mathcal{G}\left(  A\right)  =e^{-i\theta}U\operatorname{diag}\left(
\lambda_{2},\lambda_{1}\right)  $. In particular, $\frak{D}$ preserves Schmidt coefficients.
\end{enumerate}
\end{enumerate}
\end{theorem}

For $N=2$, $\frak{D}$ is an analogue of conjugation of coordinates in the
so-called ``magic basis'' of \cite{Hill and Wootters}, which is recovered by
simply removing the bars from $\left(  \ref{wootersbar}\right)  $%
.\footnote{For $N=2$ the properties \ref{maxent prop}, \ref{preconcur prop},
\ref{conj prop}, and \ref{sep prop} are just the $\mathcal{H}\otimes
\mathcal{H}^{\ast}$ analogues of the useful properites of the magic basis.} We
note that Vollbrecht and Werner \cite{Vollbrecht and Werner} make the
following point:

\begin{quotation}
\noindent The remarkable properties of the [magic] basis...are in some sense
not so much a property of that basis, but of the antiunitary operation of
\textit{complex conjugation} in [that] basis.
\end{quotation}

\noindent In particular, one may canonically translate the results and the
magic-basis or magic-conjugation proofs of \cite{Hill and Wootters}%
\cite{Wootters}\cite{makhlin}\cite{Kraus and Cirac} on $\mathbb{C}^{2}%
\otimes\mathbb{C}^{2}$ into basis-free results and proofs on $\mathcal{H}%
\otimes\mathcal{H}^{\ast}$. Hence it is apparent that choice of a basis (or of
a basis-dependent conjugation) is necessary in the cited proofs because choice
is necessary to select an isomorphism between $\mathcal{H}\otimes\mathcal{H} $
and $\mathcal{H}\otimes\mathcal{H}^{\ast}$.

\begin{acknowledgement}
We would like to thank Mike Nielsen, Andreas Klappenecker, Bahman Saffari,
Harold Shapiro, and Petre Dita for their correspondence, J. I. Cirac and G.
Vidal for pointing out their interesting reference, and Arthur Jaffe for his encouragement.\newpage
\end{acknowledgement}

\end{document}